\documentclass[a4paper]{article}
\usepackage{multirow}
\usepackage{INTERSPEECH2019}

\def\z{{\mathbf{z}}}
\def\C{{\mathbf{C}}}

\title{The DKU-DukeECE Systems for \\ VoxCeleb Speaker Recognition Challenge 2020}
\name{  Weiqing Wang$^1$, Danwei Cai$^1$, Xiaoyi Qin$^2$, Ming Li$^{1,2}$}

\address{
$^1$Department of Electrical and Computer Engineering, Duke University, Durham, USA \\ 
$^2$Data Science Research Center, Duke Kunshan University, Kunshan, China
}
\email{ming.li369@duke.edu}

\begin{document}

\maketitle
\begin{abstract}
In this paper, we present the system submission for the VoxCeleb Speaker Recognition Challenge 2020 (VoxSRC-20) by the DKU-DukeECE team. For track 1, we explore various kinds of state-of-the-art front-end extractors with different pooling layers and objective loss functions. For track 3, we employ an iterative framework for self-supervised speaker representation learning based on a deep neural network (DNN). For track 4, we investigate the whole system pipeline for speaker diarization, including voice activity detection (VAD), uniform segmentation, speaker embedding extraction, and clustering.
\end{abstract}

\section{System Descriptions for Track 1}
\begin{table*}[tp]
  \caption{ The performance of different speaker verification system. The SN denotes the score normalization}

  \label{tab:task1_result}
  \centering
  \begin{tabular}[c]{llllllllllll}
    \toprule
     \multirow{2}*{\textbf{Model}} & \multicolumn{2}{c}{\textbf{VoxCeleb1}} &  \multicolumn{2}{c}{\textbf{VoxCeleb1-E}} & \multicolumn{2}{c}{\textbf{VoxCeleb1-H}} & \multicolumn{2}{c}{\textbf{VoxSRC20-dev (SN)}} \\
     \cmidrule(lr){2-3} \cmidrule(lr){4-5} \cmidrule(lr){6-7} \cmidrule(lr){8-9} 
     
      & \textbf{EER[\%]} & \textbf{mDCF}$_{0.01}$ & \textbf{EER[\%]}  & \textbf{mDCF}$_{0.01}$ & \textbf{EER[\%]}  & \textbf{mDCF}$_{0.01}$ & \textbf{EER[\%]}  & \textbf{mDCF}$_{0.05}$ &\\
 
   \midrule
   
	1 ResNet & 0.888 & 0.088 & 1.133 & 0.124 & 2.008 & 0.199 & 2.988 & 0.1584  \\
	
	2 ResNet-BAM & 0.941 & 0.088 & - & - & - & -  & 3.020 & 0.1537  \\
	
	3 ECAPA-TDNN & 0.792 & 0.084 & 1.042 & 0.116 & 1.959 & 0.194 & 3.094 & 0.1599  \\
	 \midrule
	 Fusion(1+2+3) & - & - & - & - & - & - & 2.4798 & 0.1252\\
	  
     \bottomrule
     \end{tabular}
\end{table*}
\subsection{Data}
The experiments are conducted on the development set of Voxceleb 2, which contains 1,092,009 utterances from 5,994 speakers \cite{chung_voxceleb2:_2018}. 
For evaluation, the development set and test set of Voxceleb 1 are used \cite{nagrani_voxceleb:_2017}. We report the experimental results on 3 trial sets as defined in \cite{chung_voxceleb2:_2018}: the \textit{original test set} of Voxceleb 1 containing 37,720 trials from 40 speakers, the \textit{Voxceleb 1-E} test set (using the entire dataset) containing 581,480 trials from 1251 speakers, the \textit{Voxceleb 1-H} test set (within the same nationality and gender) containing 552,536 trials from 1190 speakers.

\subsection{Data Augmentation}

We perform online data augmentation \cite{cai_--fly_2020} with MUSAN dataset \cite{musan}. The noise type includes ambient noise, music, television, and babble noise for the background additive noise. The television noise is generated with one music file and one speech file. The babble noise is constructed by mixing three to eight speech files into one. For the reverberation, the convolution operation is performed with 40,000 simulated room impulse responses (RIR) in MUSAN. We only use RIRs from small and medium rooms. 

Moreover, we adopt the speed perturbation using sox to increase the speaker number. The strategy also has a successful application in speech and speaker recognition tasks \cite{speed_perturb_spk,speed_perturb_speech}. We speed up or down each utterance by 0.9 or 1.1 times, and the utterances with different speeds are considered from new speakers. Finally, we have 1,092,009$\times$3 = 3,276,027 utterances from 5,994$\times$3 = 17,982 speakers.

\subsection{Deep Speaker Embedding Model}

In this part, we introduce three different speaker verification systems, including the ResNet, the ResNet-BAM, and the ECAPA-TDNN. The acoustic features are 80-dimensional log Mel-filterbank energies with a frame length of 25ms and hop size of 10ms. The extracted features are mean-normalized before feeding into the deep speaker network.

\subsubsection{ResNet}
For the ResNet module, we adopt the same structure as \cite{cai_exploring_2018}. The network structure contains three main components: a front-end pattern extractor, an encoder layer, and a back-end classifier. The ResNet34 \cite{He2016Deep} structure is employed as the front-end pattern extractor, which learns a frame-level representation from the input acoustic feature. The widths (number of channels) of the residual blocks are \{32, 64, 128, 256\}. The global statistic pooling (GSP) layer, which computes the mean and standard deviation of the output feature maps, can project the variable length input to the fixed-length vector. The 128-dimensional fully connected layer following the pooling layer is adopted as the speaker embedding layer. The ArcFace \cite{arcface} (s=32,m=0.2) which could increase intra-speaker distances while ensuring inter-speaker compactness is used as a classifier . The detailed configuration of the neural network is the same as \cite{ffsvc20}. 

\subsubsection{ResNet with BAM}

Inspired by \cite{resnetbam}, we adopt the Bottleneck Attention Module (BAM) in the speaker verification system. The BAM module is placed after each basic block of the ResNet. Different from the ResNet system, we increase the widths from \{32, 64, 128, 256\} to \{64, 128, 256, 512\}. The pooling layer and the classifier is the same as the ResNet System in Section 1.3.1.

\subsubsection{ECAPA-TDNN}

The ECAPA-TDNN Network \cite{ecapa} achieves great success in the speaker verification task and provides the start-of-the-art performance. For this model, 1024 feature channels are used to scale up the network. The dimension of the bottleneck in the SE-Block is set to 256. The front-end feature extractor is followed by an attentive statistics pooling (ASP) layer \cite{asp_pooling} that calculates the mean and standard deviations of the final frame-level features. The classifier is the same as the ResNet system in Section 1.3.1.

\subsection{Score Normalization and System Fusion}

The cosine similarity is the back-end scoring method. After scoring, results from all trials are subject to score normalization. We utilize Adaptive Symmetric Score Normalization (AS-Norm) in our systems. The adaptive cohort for the enrollment file is the X closest (most positive scores) files to the enrollment utterance. The imposter cohort consists of the average of the length normalized utterance-based embeddings of each training speaker. 

In the system fusion stage, we adopt manual calibration and automatic calibration. According to the system performance in the development data set, we adopt the score level fusion that assigns weights to different models. The weights of ResNet, ResNet-BAM and ECAPA-TDNN models are $\{1,1.2,1\}$, respectively. Considering that the model may overfit on the development set with manual calibration, we use the BOSARIS Toolkit \cite{bosaris} for calibrating.

\subsection{Experimental Results}
Table \ref{tab:task1_result} shows the performance of different speaker verification system. The best system achieved 3.9620\% in EER and 0.2041 in mDCF$_{0.05}$.

\section{System Descriptions for Track 3}

\begin{table*}[htb]
  \caption{Speaker verification performance (minDCF and EER[\%]). The utterance and speaker number of the training data are presented. }
  \label{tab:results}
  \centering
  \begin{tabular}[c]{lcccc|cc|cc|cc}
    \toprule
    \textbf{Model} & \textbf{\#Utterances} & \textbf{\#Clusters} & \multicolumn{2}{c}{\textbf{Voxceleb 1 test}} & \multicolumn{2}{c}{\textbf{Voxceleb 1-E}} & \multicolumn{2}{c}{\textbf{Voxceleb 1-H}} & \multicolumn{2}{c}{\textbf{VoxSRC20-dev}} \\
    \midrule
    Initial round (CSL) & 1,092,009 & - & 0.508 & 8.86 & 0.570 & 10.15 & 0.710 & 16.20 & 0.857 & 20.11 \\
    Round 1 & 347,625 & 2,839 & 0.429 & 6.96 & 0.433 & 7.91 & 0.561 & 11.73 & 0.700 & 14.59 \\
    Round 2 & 631,408 & 4,776 & 0.341 & 5.42 & 0.358 & 6.22 & 0.479 & 9.60 & 0.606 & 12.05 \\
    \bottomrule
  \end{tabular}
\end{table*}

The framework starts with training a speaker embedding network using a contrastive self-supervised learning algorithm.
Then we cluster the speaker embeddings obtained from the previous speaker network and use the subsequent class assignments as pseudo labels to train a new DNN.
Moreover, we iteratively train the speaker network with pseudo labels generated from the previous step to bootstrap the discriminative power of a DNN.

The proposed iterative framework for self-supervised speaker embedding learning is: 
\begin{itemize}
    \item Step 1 (initial round): Train a speaker embedding network with contrastive self-supervised learning.
    \item Step 2:  With the previous speaker embedding network, extract speaker embeddings for the whole training data. Perform a clustering algorithm on the embeddings to generate pseudo labels.
    \item Step 3: Train the speaker embedding network with a classification layer and cross-entropy loss using the generated pseudo labels.
    \item Repeat step 2 and step 3 with limited rounds. Use the last speaker embedding network as the final model.
\end{itemize}

\subsection{Data and Augmentation Strategy}
The dataset is the same as that in Section 1.1, but the speaker labels are not used in the proposed method. The data augmentation strategy is similar to the description in Section 1.2.

With contrastive self-supervised learning, three augmentation types are randomly applied to each training utterance: applying only noise addition, applying only reverberation, and applying both noise and reverberation. The signal-to-noise ratios (SNR) are set between 5 to 20 dB.

When training with pseudo labels, either background noise or reverberation noise is added to the clean utterances with a probability of 0.6. The SNR is randomly set between 0 to 20 dB. 

\subsection{Contrastive Self-supervised Learning}
First, we train a speaker embedding network with contrastive self-supervised learning (CSL) similar to the framework in \cite{chen_simple_2020, falcon_framework_2020}. For feature extraction, we choose a 40-dimensional log Mel-spectrogram with a 25ms Hamming window and 10ms shifts. The duration between 2 to 4 seconds is randomly generated for each data batch.
We use the same network architecture as in \cite{cai_within-sample_2020}. ReLU non-linear activation and batch normalization are applied to each convolutional layer in ResNet. Network parameters are updated using Adam optimizer \cite{kingma_adam_2017} with an initial learning rate of 0.001 and a batch size of 256. The temperature $\tau$ is set as 0.1.

\subsection{Generating Pseudo Labels by Clustering}

\subsubsection{\textit{k}-means clustering}
Given the speaker embeddings of the training data, we employ the \textit{k}-means algorithm to generate cluster assignments. The cluster number is set to 6,000 for \textit{k}-means. In Voxceleb 2, the audio segments are obtained with the self-supervised SyncNet \cite{chung_voxceleb2:_2018} from videos. We take advantage of this segment information and average the speaker embeddings from the same video. The \textit{k}-means clustering is performed on the averaged speaker embeddings for the sake of clustering efficiency.

\subsubsection{Purifying pseudo labels}
Since the pseudo labels contain massive label noise, we apply the following simple steps to purify the generated pseudo labels.
(\textit{a}) By defining the clustering confidence as $\|\z_i-\C_{y_i}\|_2^2$ for each speaker embedding $\z_i$, we filter out $p$ portion of the remaining data with least clustering confidence.
(\textit{b}) To further reduce the possibility that one actual speaker appears in several pseudo clusters, we only keep the pseudo clusters with at least $S$ samples. The p and S are set to 0.6 and 8 for the clustered embeddings of CSL. 

\subsection{Learning with Pseudo Labels}
After obtaining the purified pseudo labels, we train the speaker embedding network with a classification layer and cross-entropy loss using the generated pseudo labels. For the network learned with pseudo labels, we use an 80-dimensional log Mel-spectrogram with a 25ms Hamming window and 10ms shifts as input features. A duration between 3 to 4 seconds is randomly generated for each data batch. The network architecture is the same as the one used in CSL but with double feature map channels. Dropout is added before the speaker classification layer to prevent overfitting \cite{srivastava_dropout:_2014}. Network parameters are updated using the stochastic gradient descent (SGD) algorithm. The learning rate is initially set to 0.1 and is divided by 10 whenever the training loss reaches a plateau. The p and S are set to 0.4 and 10 for the clustered embeddings when training this network.

\subsection{Experimental Results}
Cosine similarity is used for scoring at the test stage. We use an equal error rate (EER) and minimum detection cost function (minDCF) as the performance metric. The reported minDCF  is evaluated with $P_{\mathrm{target}}=0.05, C_\mathrm{miss}=C_\mathrm{fa}=1$. Table \ref{tab:results} reports the experimental results. In round 1, with only 32\% of the training data and pseudo labels containing label noise, the speaker model outperforms the model trained with contrastive self-supervision loss by 21.4\% in terms of EER.

\section{System Descriptions for Track 4}
Our diarization system of each recording consists of the following modules:
\begin{itemize}
    \item Voice activity detection (VAD): VAD detects speech in the recordings and removes the non-speech regions.
    \item Segmentation: The speech regions in the recordings are uniformly split into sub-segments every 0.75s from windows of 1.5s.
    \item Speaker embedding extraction: After segmentation, short segments are mapped into the speaker subspace and generate fixed-dimensional speaker embeddings.
    \item Similarity measurement: Similarity scores between any two speaker embeddings in the same recording are computed and used for the clustering step.
    \item Clustering: Clustering algorithms assign segments with high similarity scores to the same cluster and finally give the diarization results. 
    \item Resegmentation: Resegmentation modules can estimate distributions of clustering results and refine them by frames.
    \item Overlap detection: Overlap detection can further refine the diarization results by estimating overlapped speech frames. 
    
\end{itemize}
\subsection{Data}
For speaker embedding training, we use Voxceleb1\&2 \cite{voxceleb}  for speaker embedding training. For VAD, overlap detection, and speaker diarization training, several 16k sampled dataset are used, including AMI \cite{ami}, ICSI \cite{icsi}, ISL (LDC2004S05), NIST (LDC2004S09) and SPINE1{\&}2 (LDC2000S87, LDC2000S96, LDC2001S04, LDC2001S06, LDC2001S08). MUSAN \cite{musan} and RIRS corpora are employed for data augmentation for all training steps, which is the same as the augmentation method in Section 1.2.

For validation on the Voxconverse dev set \cite{voxconverse}, we regard it as a held-out dataset. It means neither threshold tuning nor model adaptation is performed. As for evaluation on the Voxconverse test set, we take the first 150 recordings in the dev set as the finetuning set to finetune our models and improve the performance, and then validate on the remaining 60 recordings in the dev set to avoid overfitting.

\subsection{VAD}
We use a ResNet-based network for the VAD task. The structure of the network is the same as the model in \cite{vad1, vad2}, which consists of a front-end ResNet and a back-end classifier with the Bi-LSTM and fully-connected layers. 

To be specific, we extract a 64-dimensional log Mel-spectrogram with a 25ms Hamming window and 10ms shifts as input features. An input sequence contains 800 frames. ResNet34 is employed and its channel width is set to \{32, 64, 128, 256\}. Two Bi-LSTM layers are stacked, each with 64 units per direction and a dropout rate of 0.5, followed by two linear layers with 64 units and 1 unit, respectively.  Training sets include 16k meeting data with data augmentation, as described in Section 3.1.  The stochastic gradient descent (SGD) optimizer and the binary cross-entropy (BCE) loss are employed. The learning rate is initialized as 0.1 and decreases by a factor of 0.1 whenever the training loss reaches a plateau. For evaluation, we take the first 150 recordings in the dev set to finetune the model and use the model with the lowest loss on the remaining dev set as the final model.

\subsection{Segmentation}
We use the uniform segmentation with overlap rather than speaker change detection (SCD). In our experiments, the sliding window is 1.5s with 0.75s shift. The ground-truth speaker of a sub-segment is the one who talks most in the central 0.75s region.

\subsection{Speaker Embedding Training and Extraction}
We use the Deep ResNet as the speaker embedding network, and the model configuration is the same as the ResNet described in Section 1.3.1 for track 1. The only difference is that the input duration is 1 to 2 seconds, and the loss function is the softmax. 

In our experiments, the training sets include voxceleb1\&2 with data augmentation, where there are 7323 speakers in total. The SGD optimizer and cross-entropy loss are employed. The initial learning rate is 0.1 and reduces whenever the training loss reaches a plateau. For testing cases, outputs from the first linear layer are taken as 128-dimensional speaker embeddings.

\subsection{Similarity Measurement}
We employ the self-attention mechanism in the similarity measurement, and the framework of this module is similar to the Attentive vector-to-sequence (Att-v2s) described in \cite{linself}. In the Att-v2s model, the first linear layer is 256-dimensional. The encoder layer contains 2 heads with 128 attention units for each head, and the dimension of the feed-forward layer is 1024. The second linear is 1-dimensional, connected with the Sigmoid
function. 

The input data is the speaker embeddings from each sub-segments. In the data preparation stage, thousands of 1.5s speaker embeddings are extracted from each recording. Then during the training stage, we randomly truncate 200 to 400 successive speaker embeddings as the input sequence. The output is one row of the affinity matrix $\mathbf{S}$. For the i-th row $\mathbf{S}_i$, it can be calculated as follows:

\begin{equation}
    \bm{S}_{i} = [S_{i1}, S_{i2}, ..., S_{in}] = f_{\theta}(\bm{m}_{i})\\
\end{equation}
\begin{equation}
    \bm{m}_{i} = \left[\begin{matrix} \bm{x}_i & \bm{x}_i & ... & \bm{x}_i\\\bm{x}_1 &\bm{x}_2 & ... & \bm{x}_n\end{matrix}\right], 
\end{equation}
where $f$ is the neural network, $\theta$ is the parameter of the network, $\bm{m}_{i}$ is the input of the network, and $\bm{x}_{i}$ is the i-th speaker embedding in a sequence.

Training features are speaker embeddings extracted from 16k meeting data described in Section 3.1. The SGD optimizer and the BCE loss are employed. The learning rate is initialized as 0.01 and decays by a factor of 0.1 whenever the training loss reaches a plateau. The training process terminates after 100 epochs, and the model is validated on the dev set. For evaluation, we further take the first 150 recordings in dev set to finetune our model and test on the remaining 66 recordings.

\subsection{Clustering}
In the experiments, we employ spectral clustering \cite{spectral-clustering} as the back-end clustering method. The affinity matrices are first symmetrized and normalized, as mentioned in \cite{SD-lstm}. Then we estimate the number of the speaker with a threshold of 0.99 and finally get the diarization results.

\subsection{Resegmentation}
In our experiments, we employ the VB diarization method and the tool released by \cite{reseg}. UBM with 1024 components are trained on 24-dimensional MFCCs and z is 400-dimensional. Other parameters for the VB algorithm are: maxIters=1, downsample=3, loopProb=0.99, statScale=0.3.

\subsection{Overlap Detection}
The model structure, data, and training configurations are all the same as those in Section 3.2 of VAD. The only difference is that the label is 1 for overlapped regions and 0 otherwise. For testing cases, when a segment is classified as overlapped speech in audios, we extend its boundary by ±15 frames and take all speakers appearing in the extended segment as labels of the original segment.

\subsection{Experimental Results}
Both DER and JER with a collar of 0.25 are reported in Table \ref{tab:track4}. We first directly test on the all dev set (dev-all), then test on the last 66 recordings in the dev set (dev-66) without any threshold tuning or model adaptation. Then we finetune the VAD, similarity measurement, and overlap detection model on the first 150 recordings in the dev set and test on dev-66.

\begin{table}[htb]
  \caption{JER and DER on voxconverse dev set}
  \label{tab:track4}
  \centering
  \begin{tabular}[c]{lc|cc}
    \toprule
    \textbf{dataset} & \textbf{+finetune} & \textbf{DER} & \textbf{JER}  \\
    \midrule
    dev-all & no & 9.74 & 36.08 \\
    dev-66 & no & 7.53 & 35.81 \\
    dev-66 & yes & 6.46 & 28.58 \\
    \bottomrule
  \end{tabular}
\end{table}


\bibliographystyle{IEEEtran}

\bibliography{mybib}


\end{document}